\documentclass[pra,aps,twocolumn,superscriptaddress,showpacs,floatfix]{revtex4}

\usepackage{simon}
\usepackage{graphicx}
\usepackage[letterpaper,margin=1in]{geometry}

\begin{document}


\title{A relation between electromagnetically induced absorption resonances and nonlinear magneto-optics in $\Lambda$-systems}
\author{D. Budker}\email{budker@socrates.berkeley.edu}
\affiliation{Department of Physics, University of California at
Berkeley, Berkeley, California 94720-7300} \affiliation{Nuclear
Science Division, Lawrence Berkeley National Laboratory, Berkeley,
California 94720}
\author{S. M. Rochester}\email{simonkeys@yahoo.com}
\affiliation{Department of Physics, University of California at
Berkeley, Berkeley, California 94720-7300}

\date{\today}

\begin{abstract}
Recent work on $\GL$-resonances in alkali metal vapors (E.
Mikhailov, I. Novikova, Yu.\ V. Rostovtsev, and G. R. Welch,
quant-ph/0309171, and references therein) investigated a type of
electromagnetically induced absorption resonance that occurs in
three-level systems under specific conditions normally associated
with electromagnetically induced transparency. In this note, we
show that these resonances have a direct analog in nonlinear
magneto-optics, and support this conclusion with a calculation for
a $J=1\rightarrow J'=0$ system interacting with a single nearly
circularly polarized light field in the presence of a weak
longitudinal magnetic field.
\end{abstract}



\maketitle

Electromagnetically induced transparency (EIT, see, for example,
Ref.\ \cite{Har97} and references therein) is a phenomenon in
which absorption of a light field by a resonant system is
inhibited by the presence of an additional electromagnetic field.
There is a close connection, discussed in detail in Ref.\
\cite{Bud2002RMP}, between EIT and resonant nonlinear magneto- and
electro-optical effects (NMOE), e.g., nonlinear Faraday rotation.
Recently, a counterpart to EIT, electromagnetically induced
\textit{absorption} (EIA), has also garnered significant
experimental and theoretical attention (see, for example, Refs.\
\cite{Aku98,Aln2003,Fai2003} and references therein). This effect
is also closely related to NMOE. For example, consider optical
pumping on a $F\rightarrow F'=F+1$ transition, a situation in
which EIA may occur \cite{Kaz84}. In studies of nonlinear
magneto-optical rotation (NMOR) on the Rb $D2$-line, the sign of
optical rotation was seen to change as the light frequency is
scanned across an $F\ra F'$ transition group [see Fig.\ 15(a) in
the review paper \cite{Bud2002RMP}]. This dependence results from
the fact that, because of EIA, rotation due to a $F\ra F+1$
transition is opposite in sign to that due to $F\ra F'=F-1,F$
transitions, for which EIA does not occur
\cite{Kan93,Bud99,Bud2002RMP}.

\begin{figure}
    \includegraphics[width=3.25in]{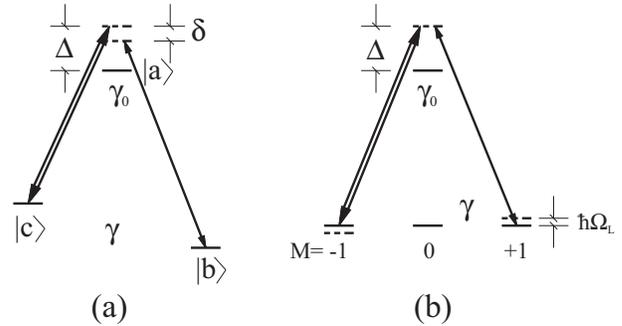}
    \caption{(a) A nondegenerate $\Lambda$-system subject to a
    bichromatic phase-coherent light field. (b) A $J=1\rightarrow
    J'=0$ atomic transition subject to a monochromatic
    elliptically-polarized light field and a longitudinal magnetic
    field. $\Delta$ is the one-photon detuning, $\gamma_0$ is the
    homogeneous width of the optical transition, $\gamma$ is the
    ground-state relaxation rate, $\Gd$ and $\Omega_L$ are the
    two-photon detuning and analogous Larmor frequency,
    respectively.} \label{Fig:System}
\end{figure}

A type of EIA resonance exhibited by a three-level $\GL$-system
[Fig.\ \ref{Fig:System}(a)] interacting with a bichromatic light
field was recently investigated  both experimentally and
theoretically \cite{Mik2003a,Mik2003b}. Experiments were carried
out with $^{87}$Rb atoms contained in vapor cells with and without
buffer gas.  The two ground-state hyperfine components of
$^{87}$Rb were used as the two lower states $\ket{b}$ and
$\ket{c}$. Two phase-coherent copropagating laser fields tuned
near the $D1$ transition were used as the drive and the probe
fields.
The intensity of the probe field was analyzed at the output of the
cell as a function of the two-photon (Raman) detuning $\Gd$. For a
buffer-gas-free cell, EIT, manifesting itself as a peak in the
output probe-field intensity at near-zero two-photon detuning, was
observed at all one-photon detunings $\GD$. However, for cells
with buffer gas (for example, 30 torr of Ne), very different
behavior was seen. For small one-photon detuning $\GD$, EIT was
observed, as in the buffer-gas-free case. At larger values of
$\GD$, the resonance in two-photon detuning became asymmetric,
eventually turning into an EIA resonance. A rather striking
feature of the EIA resonance is that it leads to a significant
reduction of the transmitted probe-field intensity on the
two-photon resonance under conditions where there is very little
absorption of the probe field in the absence of the drive. The
authors of Refs.\ \cite{Mik2003a,Mik2003b} found that, for the EIA
effect to occur, the buffer-gas collisions must broaden the
optical transition to the point that the homogeneous width is
comparable to or exceeds the Doppler width.

The EIT resonance observed for small one-photon detuning is the
standard coherent population trapping effect: at zero two-photon
detuning the drive and probe fields combine coherently to
optically pump the atoms into a state that does not couple to the
combined field. The EIA resonance, in contrast, is observed when
the light is far detuned from the one-photon resonance. Under
two-photon (Raman) resonance conditions (when the two-photon
detuning equals the differential ac Stark shift), atoms are
transferred from the state that interacts with the drive light to
the state that interacts with the probe light. This causes
increased absorption on the probe transition \cite{Lou92,
Bra2004}.

In this note, we point out the connection between these EIA
resonances and a nonlinear magneto-optical effect arising when a
single elliptically polarized light field interacts with atoms in
the presence of a weak magnetic field applied along the light
propagation direction. We consider a $J=1\rightarrow J'=0$
transition [Fig.\ \ref{Fig:System}(b)] and associate the states
$\ket{b}$ and $\ket{c}$ with the $M=-1$ and $M=1$ Zeeman sublevels
of the $J=1$ ground state, and the two components of the incident
light field with the left- and right-circular polarizations of a
monochromatic light field with wavelength $\Gl$. The resultant
light field corresponds to elliptically polarized light. Since one
circular component, say the right circular one, is much weaker
than the other, the light deviates only slightly from pure
circular polarization. In other words, the angle of ellipticity
$\Ge$, equal to the arctangent of the ratio of the minor to the
major axis of the polarization ellipse, is close to $\Gp/4$. The
relative phase of the two circular components, which defines the
angle of orientation $\Gv$ of the major axis of the polarization
ellipse, can have any value. In order to complete the analogy
between the two problems, we need to introduce the analog of the
two-photon detuning $\delta$. This is done by applying a
longitudinal magnetic field that splits the Zeeman sublevels by
the Larmor frequency $\GO_L$ [Fig.\ \ref{Fig:System}(b)].

Studies of magneto-optical effects are commonly done by measuring
output polarization rather than transmission of a field component.
To relate the two in this case, we write the total complex optical
field $\wt{\mb{E}}$ (assumed to be propagating in the $\uv{z}$
direction) as
\begin{equation}
\begin{split}
    \wt{\mb{E}}
    =E_0&\left[\prn{\cos\Ge\cos\Gv-i\sin\Ge\sin\Gv}\uv{x}\right.\\
    &\left.{}+\prn{\cos\Ge\sin\Gv+i\sin\Ge\cos\Gv}\uv{y}\right]e^{-i\prn{\Go t-kz-\Gf}},
\end{split}
\end{equation}
where $E_0$ is the electric field amplitude and $\Gf$ is the
overall phase. In the spherical tensor basis,
$\uv{e}_\pm=\mp\prn{\uv{x}\pm i\uv{y}}/\sqrt{2}$,
\begin{equation}
\begin{split}
    \wt{\mb{E}}
    ={}&-E_0\prn{e^{i\Gv}\sin\Ge'\uv{e}_-
        +e^{-i\Gv}\cos\Ge'\uv{e}_+}e^{-i\prn{\Go t-kz-\Gf}}\\
    ={}&\prn{E_-\uv{e}_-+E_+\uv{e}_+}
        e^{-i\prn{\Go t-kz}},
\end{split}
\end{equation}
where we have defined $\Ge'=\Ge-\Gp/4$. The intensity $I_-$ of the
right circular component is given by
\begin{equation}
    I_-
    =\frac{c}{8\Gp}E_-E_-^*
    =\frac{c}{8\Gp}E_0^2\sin^2\Ge',
\end{equation}
and a differential change in intensity over a distance $dz$ is
given by
\begin{equation}
\label{eq:dI}
    \frac{1}{I_-}\frac{\dd{I_-}}{\dd{z}}
    =\frac{2}{\Ge'}\frac{\dd\Ge'}{\dd{z}}+\frac{2}{E_0}\frac{\dd{E_0}}{\dd{z}},
\end{equation}
for small $\Ge'$. Under conditions of saturated absorption, such
as considered here, the term containing $\dd{E_0}/\dd{z}$ in Eq.\
(\ref{eq:dI}) can be neglected. The original problem is thus
analogous to the determination of the magnetically induced change
of ellipticity of nearly circularly polarized input light.

The effect on the input light polarization can be found using
standard density matrix techniques (a description of our approach
is given in Ref.\ \cite{Bud2002RMP}). In order to reproduce the
basic phenomenon---the change from EIT to EIA with the increase of
the one-photon detuning---it is sufficient to consider the
Doppler-free case. In fact, as pointed out in Ref.\
\cite{Mik2003b} and as discussed below, Doppler broadening leads
to a suppression of the effect of interest. (This is the reason
that buffer gas is required to observe the effect in the
Doppler-broadened case: by increasing the homogenous width of the
optical transition so it becomes comparable to the Doppler width,
it effectively ``turns off'' the suppression.)

The atomic Hamiltonian is
\begin{equation}
    H\simeq
        \begin{pmatrix}
            -\GO_L  &0  &0          &-\GO_R    \\
            0       &0  &0          &0         \\
            0       &0  &\GO_L      &-\Ge'\GO_R\\
            -\GO_R  &0  &-\Ge'\GO_R &-\GD
        \end{pmatrix},
\end{equation}
under the rotating wave approximation, where $\GO_L$ is the Larmor
frequency (Fig.\ \ref{Fig:System}), and
$\GO_R=||d||E_0/(2\sqrt{6})$ is the Rabi frequency, where $||d||$
is the reduced dipole matrix element. We find the steady-state
density matrix for a medium of atomic density $N_0$ evolving
according to $H$, assuming radiative decay of the upper state to
the ground state at rate $\Gg_0$, and much slower relaxation and
incoherent repopulation of the ground state, due to atoms entering
and leaving the interaction volume, at rate $\Gg$. From this, the
change in light ellipticity per unit distance through the medium
can be obtained:
\begin{equation}\label{Eq:deFormula}
    \frac{1}{\Ge'}\frac{\dd\Ge'}{\dd{z}}
    =-\frac{x\prn{D+x}\Gk N_0\Gl^2}
        {2\Gp\prn{3D^2+2\Gk}\sbr{D^2+x^2+\prn{\Gk-2Dx}^2}},
\end{equation}
where $x=\GO_L/\Gg$ is the normalized Larmor frequency,
$D=\GD/\Gg_0$ is the normalized light detuning, and the optical
pumping saturation parameter $\Gk=\GO_R^2/(\Gg\Gg_0)$ is assumed
to be much greater than unity. This formula is analogous to Eqs.\
(15-17) of Ref.\ \cite{Mik2003b}.

\begin{figure}
    \includegraphics{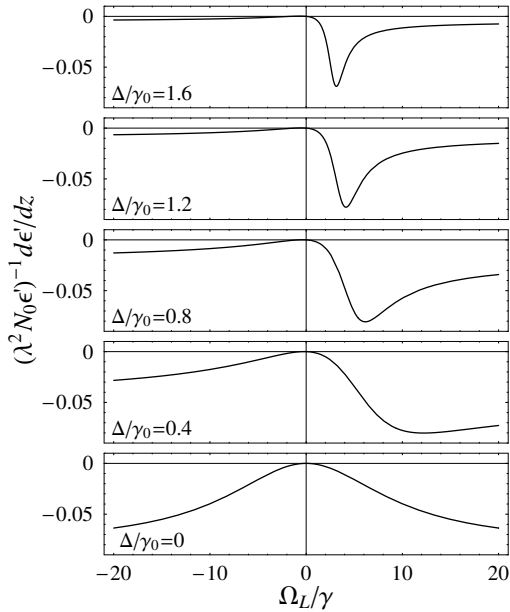}
    \caption{Change in ellipticity [Eq.\ (\ref{Eq:deFormula})] of
    almost completely circularly polarized ($\abs{\Ge'}\ll\Gp/4$)
    light as a function of normalized Larmor frequency
    $x=\Omega_L/\gamma$ at various normalized light detunings
    $D=\Delta/\gamma_0$. The value of the saturation parameter is
    $\Gk=10$ and we assume $\Gg\ll\Gg_0$. In the bottom plot, at
    zero detuning, the EIT feature is seen inside the broad
    absorption line that would exist in the absence of the drive
    field. As the detuning is increased, the EIA feature appears.}
    \label{Fig:Results}
\end{figure}

Figure \ref{Fig:Results} shows plots of the change in ellipticity
due to an optically thin sample as a function of the applied
magnetic field (the analog of the two-photon detuning $\delta$)
for different values of the one-photon detuning $\Delta$, with
$\Gk=10$.
The results shown in this figure are very similar to those
presented in Fig.\ 3 of Ref.\ \cite{Mik2003b} for the
$\Lambda$-resonances with a bichromatic light field (a decrease in
ellipticity here corresponds to absorption of probe light in the
case of Ref.\ \cite{Mik2003b}). This figure shows the change in
the sign of the two-photon resonance as the one-photon detuning is
increased (analogous to the transition from EIT to EIA). As in the
experiment presented in Ref.\ \cite{Mik2003b}, the EIA resonance
peak is shifted with respect to the EIT peak because of the
ac-Stark shift produced at nonzero one-photon detuning. Also, the
EIT peak is broader than the EIA peak due to larger power
broadening at one-photon resonance.

Figure \ref{Fig:Results} illustrates why Doppler broadening
suppresses the EIA effect: since the nonzero magnetic field
strength (two-photon detuning) at which the EIA feature occurs
depends on the light (one-photon) detuning, the feature is washed
out by Doppler averaging. Note that in Fig.\ \ref{Fig:Results} the
change in ellipticity is always zero at zero magnetic field. This
is because of the fact that, assuming ground-state relaxation due
to the transit of atoms through the light beam, as we have here, a
$1\rightarrow 0$ transition has the unusual property of producing
no change in light polarization in the absence of a magnetic field
\cite{Ale69,Roc2001SR}. (This is the same reason that, in a
standard EIT experiment, the ``dark resonance'' always occurs at
zero two-photon detuning no matter the relative power of the drive
and probe fields \cite{Lou92}.)

The analogy presented here underscores the fact that despite the
remarkable feature of the EIA resonances observed in Refs.\
\cite{Mik2003a,Mik2003b}, namely that there is significant
reduction of the transmitted probe-field intensity on the
two-photon resonance under conditions where there is very little
absorption of the probe field in the absence of the drive, the EIA
resonance is not actually associated with increased overall
absorption. Rather, a change in ellipticity is caused by the
transfer of the field between the circular polarizations. A
similar interaction between strong and weak (actually, vacuum)
polarization modes is discussed in Ref.\ \cite{Mat2002}.

We hope that the analogy between $\GL$-resonances with a
bichromatic light field and the nonlinear magneto-optical effects
will prove to be fruitful for qualitative understanding of both
phenomena and serve as a guide to further experimental and
theoretical work. Such a scenario has already been played out in
the study of light-propagation dynamics, in which a similar
analogy \cite{BudGroupVel99} led to light-propagation experiments
with monochromatic light and Zeeman sublevels that were more
straightforward than their counterparts involving bichromatic
light fields and hyperfine sublevels (see Refs.\
\cite{Boy2002,Mat2001SL} for reviews).

The authors acknowledge useful discussions with Marcis Auzinsh,
Derek Kimball, Irina Novikova, Szymon Pustelny, Jason Stalnaker,
and Valeriy Yashchuk. This work has been supported by the Office
of Naval Research (grant N00014-97-1-0214).

\bibliography{NMObibl}

\end{document}